\documentclass[preprintnumbers,article,amsmath,amssymb,floatfix,10pt,prd,onecolumn,
superscriptaddress,nofootinbib]{revtex4}
\usepackage{bbm}
\usepackage{amsfonts}
\usepackage{mathrsfs}
\usepackage{latexsym}

\usepackage{epsfig}
\usepackage{epstopdf}
\usepackage{placeins}
\usepackage{float}
\usepackage{graphicx}
\usepackage{amssymb}
\usepackage{amsmath}
\usepackage{dcolumn}
\usepackage{bm}
\usepackage{color}
\usepackage{comment}
\usepackage{xcolor}

\begin{document}

\title{\bf Existence of Wormholes in $f(\mathcal{G})$ Gravity using Symmetries}

\author{Tayyaba Naz}
\email{tayyaba.naz@nu.edu.pk}\affiliation{National University of Computer and
Emerging Sciences,\\ Lahore Campus, Pakistan.}
\author{G. Mustafa}
\email{gmustafa3828@gmail.com}\affiliation{Department of Physics, Zhejiang Normal University,
Jinhua, 321004, China.}
\author{M. Farasat Shamir}
\email{farasat.shamir@nu.edu.pk}

\begin{abstract}
The current study examines the geometry of static wormholes with anisotropic matter distribution in context of modified $f(\mathcal{G})$ gravity. We consider the well known Noether and conformal symmetries, which help in investigating wormholes in $f(\mathcal{G})$ gravity. For this purpose, we develop symmetry generators associated with conserved quantities by taking into consideration the $f(\mathcal{G})$ gravity model. Moreover, we use the conservation relationship gained from the classical Noether method and conformal Killing symmetries to develop the metric potential. These symmetries provide a strong mathematical background to investigate wormhole solutions by incorporating some suitable initial conditions. The obtained conserved quantity performs a significant role in defining the essential physical characteristics of the shape-function and energy conditions. Further, we also describe the stability of obtained wormholes solutions by employing the equilibrium condition in modified $f(\mathcal{G})$ gravity. It is observed from graphical representation of obtained wormhole solutions that Noether and conformal Killing symmetries provide the results with physically accepted patterns.
\end{abstract}
\maketitle
\section{Introduction}

The approximation symmetry method played a significant role in evaluating
the precise solutions of the differential equations.
Such approximations dynamically reduce the complexity of the non-linear
equation involved in a scheme by seeking the unknown parameter of equations.
The Noether symmetries, in particular, are not only a mechanism for dealing
with the dynamics solution, but their presence also provides suitable
conditions so that one can specify the universe models physically
and analytically according to our measured observations.
In addition to this, Noether symmetry technique is believed to be a
suitable mathematical approach, which often investigates the exact solutions and computes
the associated conserved quantities. This method plays a central role in reducing the nonlinear equation system to a linear equation system.
The numerous conservation principles, such as conservation of energy and angular momentum etc., are specifically linked to the symmetries of a specified dynamic system and provide the conserved quantities, which seem to be the consequence of certain type of symmetry being present in that mechanism.
Moreover, conserved quantities can be determined by applying the Noether symmetry technique, asking for the Lagrangian symmetry.
The presence of any specific type of symmetry for the Euler-Lagrange equations of motion, along with the Lagrangian, will precisely be related to the Noether's symmetry.
Whereas, no particular theory is endorsed by the technique of Noether symmetry, the literature studies have indicated that the existence of Noether symmetries is capable of selecting suitable theory and then integrating dynamics through the first integrals referring to Noether symmetries \cite{1}-\cite{5}.
In fact, it should be noticed that the Noether symmetries are not just a mathematical method for solving or reducing dynamics, yet their presence even enables to choice of observable universes/wormholes/black holes, etc. and the collection of analytical models relevant to observations \cite{6}. Recently, we have proposed some compact star solutions incorporating Noether
symmetry in frame of the modified $f(\mathcal{G})$ gravity \cite{fartay}. The Noether Symmetry approach for $f(\mathcal{G})$ cosmology in $n$ dimensions has been discussed in \cite{BC1}. Moreover, in spherically symmetric context, the $f(\mathcal{G})$ theory of gravity can be employed to address general relativity ($\mathcal{GR}$) inconsistencies \cite{BC2}. Further, a detailed overview of the Noether symmetry technique to investigate a variety of cosmic scenarios, including viable mimetic $f(R)$ and $f(R, T)$ theories is given in \cite{Momen}-\cite{CAPO}. In this regard, this approach has successively used to cope with cosmologies generated from various theories of gravity \cite{BC3}-\cite{BC5}.

Our universe often exhibits eye-opening challenges for cosmologists, regrading their fascinating and enigmatic existence.
The presence of hypothetical geometries are perceived to be the most contentious topic leading to wormhole geometry.
A debate regarding the existence of the wormhole and the construction of its solutions is among the most interesting challenges in modern astrophysics. A wormhole is a path or tunnel that connects two separate regions of the same or two different type of universes.
Flamm \cite{7} used the term bridge for the very first time in 1916.
In 1935, Einstein and Rosen mathematically described such bridge as structures renowned as the Einstein-Rosen bridges \cite{8}.
In addition, Morris and Throne \cite{9} established wormholes by taking into account exotic matter.
Exotic matter is regarded as the necessary component for the formation of these wormholes.
The existence of exotic matter by using various techniques have been addressed by many authors \cite{10}-\cite{12}.
Moreover, extra geometric terms are thought to be the cause of these exotic matter in modified theories of gravity \cite{13}-\cite{17}.
Recently, Sharif and Nawazish \cite{Sharif1,Sharif2} have explored static wormhole solutions
utilizing the Noether symmetry methodology in modified gravity and they noticed the stable
structure of red-shift functions for various cases.
Furthermore, Sharif and Hussain \cite{Sharif3} used the same technique to
explore wormhole physical presence in frame of $f(\mathcal{G},T)$ gravity and
investigate its properties of fluid distributions for both dust and non-dust case.

The current cosmic accelerated expansion has always been considered
to be the most revolutionizing reality on the landscape of theoretical
and observational modern cosmology. In order to take into account the late-time
accelerated expansion, two main approaches have been proposed. The
first effective way to describe the idea of accelerated cosmos expansion
in context of $\mathcal{GR}$ is the existence of dark energy,
which exhibits strong negative pressure. The second innovative approach to ponder
this concept of universe expansion is to modify the Einstein-Hilbert action at large scales.
These modifications of the $\mathcal{GR}$ play an influential role in revealing the
intriguing dynamics behind the expansion of the universe.
Among the various gravitational theories, the theory that has acquired
the prominence in the last few years is modified $f(\mathcal{G})$
gravity \cite{Noj1}. This modified gravity was obtained by incorporating the function $f(\mathcal{G})$ in Einstein Hilbert action.
The Gauss-Bonnet term is of great importance as it facilitates the regularization of gravitational action and can serve to avoid ghost contributions \cite{Chiba}. It is believed that $f(\mathcal{G})$ gravity is also quite helpful in explaining late cosmic acceleration and reconstructing some form of cosmological solution. Indeed, this theory was used as a significant approach for revealing
the mystical nature of the cosmos \cite{Noj2}.

Several important studies from literature have shown that anisotropic stars can be modeled utilizing solutions that endorse a single parameter group of conformal motion. Herrera and his colleagues \cite{Her1}-\cite{Her3} were among the pioneers who provided the general treatment of the spheres that accepted a single parameter category of conformal motions. Some significant findings employing conformal Killing vectors (CKVs) have been presented in literature \cite{Rah2}-\cite{Aktas}.
Nevertheless, CKVs approach is helpful to make the governing system easier to analyze by reducing the nonlinear
structure of partial differential equations (PDE's) into the ordinary
differential equations.
In the spacetime, due to conformal symmetry, some constraints on the gravitational potential are imposed.
However, the idea of CKVs  was considered in literature to investigate the presence of spherically symmetric wormholes, as the static symmetric spacetime presents a limited category of conformal motions. Kuhfitting \cite{Kuh} has recently investigated the stable wormholes solutions through CKVs and non-commutational distribution. Moreover, Rahaman \cite{Rahaman} used the CKVs technique to construct the wormhole solutions in the context of non-commutative geometry.

Motivated by the aforementioned literature, our focus is to construct the wormhole solutions admitting symmetries in the frame of conformal motion. To the best of our understanding, no attempt has yet been made to explore the wormhole solutions using the Noether symmetry technique under conformal motion by considering the $f(\mathcal{G})= \alpha\mathcal{G}^{n}$ gravity model \cite{Cog2}, where $n=2$.
For this aim, we extended the idea of Shamir and Tayyaba \cite{fartay} and use the conservation relationship gained from Noether method by incorporating some peculiar initial conditions to construct the relation of metric potential to address the formulation of wormhole.
The manuscript is organized as follows. In section $2$, we discuss some basics formulism of $f(\mathcal{G})$ gravity in frame of anisotropic matter distributions.
In section $\textbf{3}$, the geometry of wormhole by employing the Noether and conformal motion has been discussed. Conclusive remarks are presented in Section $4$.

\section{Some Formulation of Modified $f(\mathcal{G})$ Gravity}

The action for modified $f(\mathcal{G})$ gravity is expressed as \cite{Noj1}
\begin{equation}\label{5.1}
  \mathcal{S} = \int d^4x \sqrt{-g} \Bigg[\frac{\mathcal{R}}{2\kappa^2}+f(\mathcal{G})+L_m\Bigg],
\end{equation}
here $L_m$ shows matter Lagrangian, $\mathcal{R}$ being the Ricci scalar, $\kappa^2 = {8\pi G}$ represents the coupling constant term
and $f(\mathcal{G})$ is an
arbitrary function of the Gauss-Bonnet invariant term represented as
\begin{equation}\label{5.2G}
\mathcal{G} = \mathcal{R}^2 - 4\mathcal{R}_{\mu\nu}\mathcal{R}^{\mu\nu} +\mathcal{R}_{\mu\nu\sigma\rho}\mathcal{R}^{\mu\nu\sigma\rho},
\end{equation}
here $\mathcal{R}_{\mu\nu}$ and $\mathcal{R}_{\mu\nu\rho\sigma}$ specify the Ricci and Riemann tensors, respectively.
The variation of above action with respect to metric tensor yield the following field equations
\begin{eqnarray}\label{5.3}
G_{\xi\eta}+ 8\big[\mathcal{R}_{\xi\rho\eta\sigma} + \mathcal{R}_{\rho\eta}g_{\sigma\xi} - \mathcal{R}_{\rho\sigma}g_{\eta\xi} - \mathcal{R}_{\xi\eta}g_{\sigma\rho} + \mathcal{R}_{\xi\sigma}g_{\eta\rho}
\frac{\mathcal{R}}{2}(g_{\xi\eta}g_{\sigma\rho}-g_{\xi\sigma}g_{\eta\rho})\big]\nabla^{\rho}\nabla^{\sigma}f_\mathcal{G}+(\mathcal{G}f_\mathcal{G}- f)g_{\xi\eta} =\kappa^2T_{\xi\eta}.
\end{eqnarray}
An alternate representation of above aforementioned field equations (\ref{5.3}), which are familiar with $\mathcal{GR}$ may be described as
\begin{eqnarray}\label{5.4eff}
G_{\xi\eta} =\kappa^2T_{\xi\eta}^{eff},
\end{eqnarray}
the effective stress-energy tensor $T_{\xi\eta}^{eff}$ is given by
\begin{eqnarray}\label{5.5eff}
T_{\xi\eta}^{eff}=T_{\xi\eta}- \frac{8}{\kappa^2}\big[\mathcal{R}_{\xi\rho\eta\sigma} + \mathcal{R}_{\rho\eta}g_{\sigma\xi} -\mathcal{R}_{\rho\sigma}g_{\eta\xi} - \mathcal{R}_{\xi\eta}g_{\sigma\rho} +\mathcal{R}_{\xi\sigma}g_{\eta\rho} + \frac{\mathcal{R}}{2}(g_{\xi\eta}g_{\sigma\rho}-g_{\xi\sigma}g_{\eta\rho})\big]\nabla^{\rho}\nabla^{\sigma}f_\mathcal{G}
 -(\mathcal{G}f_\mathcal{G}- f)g_{\xi\eta}.
\end{eqnarray}
We consider the static, spherically symmetric spacetime \cite{Krori}
\begin{equation}\label{5.4}
ds^{2}= e^{\nu(r)}dt^2-e^{\lambda(r)}dr^2-r^2(d\theta^2+\sin^2 \theta d\Phi^2).
\end{equation}
The source of the configuration of matter presumed in this study is anisotropic in nature, represented as
\begin{equation}\label{5.5}
\mathcal{T}_{\chi\gamma}=(\rho+p_{t})\upsilon_{\chi}\upsilon_{\gamma}-p_{t}g_{\chi\gamma}+(p_{r}-p_{t})\xi_{\chi}\xi_{\gamma},
\end{equation}
here $\rho$, $p_{r}$ and $p_{t}$ indicate energy density, radial and tangential pressures respectively.
The four velocity and radial vector are symbolized by $\upsilon_{\chi}$ and $\xi_{\chi}$ respectively, which are satisfying the following condition
\begin{equation*}
\upsilon^{\alpha}=e^{\frac{-\nu}{2}}\delta^{\alpha}_{0},~~~\upsilon^{\alpha}\upsilon_{\alpha}=1,
~~~\xi^{\alpha}=e^{\frac{-\lambda}{2}}\delta^{\alpha}_{1},~~~\xi^{\alpha}\xi_{\alpha}=-1.
\end{equation*}
 Using equations (\ref{5.5eff}), (\ref{5.4}) and (\ref{5.5}), we obtain
\begin{eqnarray}\label{10}
\rho^{eff}&&=~~\rho-8e^{-2\lambda}(f_\mathcal{GGG}\mathcal{G}'^{2}+f_\mathcal{GG}\mathcal{G}'')(\frac{e^{\lambda}-1}{r^2})
+4e^{-2\lambda}\lambda' \mathcal{G}'f_\mathcal{GG}(\frac{e^{\lambda}-3}{r^2})-(\mathcal{G}f_\mathcal{G}-f),
\\\label{11}
p_{r}^{eff}&&=~~ ~p_{r}-4e^{-2\lambda}\nu'\mathcal{G}'f_\mathcal{GG}(\frac{e^{\lambda}-3}{r^2})+
(\mathcal{G}f_\mathcal{G}-f),\quad\quad\quad\quad\quad\\
\label{12}
p_{t}^{eff}&&=~~p_{t}-\frac{4e^{-2\lambda}\nu'}{r}(f_\mathcal{GGG}\mathcal{G}'^{2}+f_\mathcal{GG}\mathcal{G}'')-
\frac{2e^{-2\lambda}{\nu'}^{2}f_\mathcal{GG}\mathcal{G}'}{r}-\frac{2e^{-2\lambda}f_\mathcal{GG}\mathcal{G}'}{r}(2\nu''-3\nu'\lambda')+
(\mathcal{G}f_\mathcal{G}-f).
\end{eqnarray}
Here $\rho$, $p_r$ and $p_t$ are usual energy density, radial pressure and transverse pressure respectively.
The Gauss-Bonnet invariant term for the spherically symmetric space time  (\ref{5.4}) appears as
\begin{equation}\label{24}
\mathcal{G} = \frac{2e^{-\lambda}}{r^2}(\nu'\lambda'+{\nu'}^{2}e^{-\lambda}-3\nu'\lambda'e^{-\lambda}-2\nu''-{\nu'}^{2}+2\nu''e^{-\lambda}).
\end{equation}. To solve the field equations (\ref{10}-\ref{12}), which are extremely nonlinear, complicated and involve many unknowns, we need some suitable mathematical method. For this purpose, we use a special class of Lie point symmetries namely Noether symmetry approach.

In the current study, we consider the following integral of motion as discussed \cite{fartay}, i.e.,
\begin{eqnarray}\label{5.22}
\mathcal{I}_{1}= e^{\frac{\nu-3\lambda}{2}}\alpha [-\mathcal{G}\{e^{2\lambda}\mathcal{G} r^3+32(e^{\lambda}-1)\nu'\}
-8(e^{\lambda}-1)(r\nu'-10)\mathcal{G}'].
\end{eqnarray}
Here, we use another interesting approach in the context of symmetries, is the use of Killing vectors \cite{K1}. It has been argued that Killing symmetries form a subalgebra of Noether symmetries. Moreover, Noether equations may be termed as generalized Killing equations for some special cases \cite{K2,K3}.
Now we discuss CKVs in to connect wormhole geometry with Noether symmetry for the metric (\ref{5.4}). In a
given space time with manifold $\mathcal{M}$, the field for conformal vectors $\gamma$ is defined as
\begin{equation}\label{30}
\mathfrak{L}_{\gamma}g_{\mu\nu}=g_{\eta\nu}\gamma^{\eta}_{;\lambda}
+g_{\mu\eta}\gamma^{\eta}_{;\nu}=\Theta_{f}(r)g_{\mu\nu},
\end{equation}
where $\mathfrak{L}$ represents the Lie derivative. The expressions
$\gamma^{\eta}$ and $\Theta_{f}(r)$ denote the conformal functions.
Among all the symmetries, conformal symmetry and Noether symmetry
can approach some results with goodness as both the symmetry yield a
more profound knowledge into the geometry of spacetime. By plugging the spacetime
from Eq. (\ref{5.4}) in Eq. (\ref{30}) the following relations can be easily obtained
\begin{eqnarray*}\nonumber
\gamma^{1}\nu^{'}(r)=\Theta_{f}(r),\quad \gamma^{1}=\frac{r\Theta_{f}(r)}{2},\quad
\gamma^{1}\lambda^{'}(r)+2\gamma^{1}_{,1}=\Theta_{f}(r).
\end{eqnarray*}
The above results further lead to
\begin{eqnarray}\label{31}
e^{\nu(r)}=\Lambda_{1}^{2}r^2,\quad\quad\quad\quad\quad
e^{\lambda(r)}=\left(\frac{\Lambda_{2}}{\Theta_{f}(r)}\right)^{2},
\end{eqnarray}
where $\Lambda_{1}$ and $\Lambda_{2}$ represent the constants of integration.

\section{Wormholes in $f(\mathcal{G})$ Gravity Admitting Noether and Conformal Symmetries}

We shall describe the wormhole solutions in the background of conformal and Noether symmetries in this section. The standard spacetime for wormhole geometry is defined as
\begin{equation}\label{32}
ds^{2}= e^{2\Omega(r)}dt^2-\left(1-\frac{\mathbb{S}_{f}(r)}{r}\right)^{-1}dr^2-r^2(d\theta^2+\sin^2 \theta d\Phi^2),
\end{equation}
where $\Omega(r)$ and $\mathbb{S}_{f}(r)$ denote the redshift function and shape-function respectively.
A certain significant criteria for wormhole physics to be satisfied by the shape function and the red-shift function are summarized here. The value of the red-shift function $\Omega(r)$ must be finite within the configuration. There is no horizon restriction on $\Omega(r)$, for the wormhole to be traversable. The appropriate radial distance by enforcing the constraint in the shape function relation $\mathbb{S}_{f}(r)$,
$\mathcal {L}(r)=\pm\int^{r}_{r_{0}}(1+\frac{\mathbb{S}_{f}(r)}{r})^{-1/2}dr$, with $r>r_{0}$, should be finite every where in the spacetime geometry. Here, the $\pm$ incorporates the two different parts of the spacetime geometry, interconnected by the wormhole configuration. The upper segment of the wormhole decreases and hits its lowest at the position of wormhole of the throat, and then rises to the lower part. The $\mathbb{S}_{f}(r)$ needs to satisfy the  inequality given by $(\mathbb{S}_{f}(r)-\mathbb{S}_{f}(r)^{'}r)/\mathbb{S}_{f}(r)^{2}>0$ and the equality $\mathbb{S}_{f}(r_{0})=r_{0}$. The $\mathbb{S}_{f}(r)$ should also satisfy the condition $\mathbb{S}_{f}^{'}(r_{0})<1$. By equating the spherically symmetric spacetime (\ref{5.4}) with Eq. (\ref{32}), we get the following relations
\begin{equation}\label{33}
g_{tt}=e^{\nu(r)}=e^{2\Omega(r)},\quad\quad\quad\quad\quad g_{rr}=e^{\lambda(r)}=\left(1-\frac{\mathbb{S}_{f}(r)}{r}\right)^{-1}
=\left(\frac{\Lambda_{2}}{\Theta_{f}(r)}\right)^{2}.
\end{equation}
Using Eq. (\ref{33}), (\ref{10}-\ref{12}) and considering $\Lambda_{1}^{2}=\Lambda_{3}$, $\Lambda_{2}^{2}=\Lambda_{4}$,  $(\Theta_{f}(r))^2=\Theta_{f_0}(r)$, we get the simplified system of field equations
\begin{eqnarray}
\rho^{eff}&&=\frac{1}{\Lambda_3^4 r^{15} \Theta_{f_{0}}(r)^4}\bigg(540 \alpha  r^3 \left(\Lambda_3 r^2-1\right){}^2 \Theta_{f_{0}}'(r)^4-16 \alpha  r^2 \Theta_{f_{0}}(r) \left(\Lambda_3 r^2-1\right) \Theta_{f_{0}}'(r)^2 \left(65 r \left(\Lambda_3 r^2-1\right) \Theta_{f_{0}}''(r)\right.\nonumber\\&&\left.+\left(195-113 \Lambda_3 r^2\right) \Theta_{f_{0}}'(r)\right)+16 \alpha  r \Theta_{f_{0}}(r)^2 \left(15 r^2 \left(\Lambda_3 r^2-1\right){}^2 \Theta_{f_{0}}''(r)^2+\left(\Lambda_3 r^2 \left(187 \Lambda_3 r^2-706\right)+555\right)\right.\nonumber\\&&\times\left. \Theta_{f_{0}}'(r)^2+2 r \left(\Lambda_3 r^2-1\right) \Theta_{f_{0}}'(r) \left(10 r \Theta_{f_{0}}^{(3)}(r) \left(\Lambda_3 r^2-1\right)+\left(135-77 \Lambda_3 r^2\right) \Theta_{f_{0}}''(r)\right)\right)+64 \alpha  \Theta_{f_{0}}(r)^3 \nonumber\\&&\times\left(\left(\Lambda_3 r^2 \left(35 \Lambda_3 r^2-234\right)+231\right) \Theta_{f_{0}}'(r)+r \left(\left(\Lambda_3 r^2 \left(138-35 \Lambda_3 r^2\right)-111\right) \Theta_{f_{0}}''(r)+r \left(\Lambda_3 r^2-1\right)\right.\right.\nonumber\\&&\times \left.\left.\left(r \Theta_{f_{0}}^{(4)}(r) \left(1-\Lambda_3 r^2\right)+2 \Theta_{f_{0}}^{(3)}(r) \left(5 \Lambda_3 r^2-9\right)\right)\right)\right)+\Lambda_3^4 \rho  r^{15} \Theta_{f_{0}}(r)^4\bigg),\label{35}\\
p^{eff}_r&&=\frac{1}{\Lambda_3^4 r^{14} \Theta_{f_{0}}(r)^4}\bigg(-12 \alpha  r^2 \left(\Lambda_3 r^2-1\right) \left(5 \Lambda_3 r^2-21\right) \Theta_{f_{0}}'(r)^4+16 \alpha  r \Theta_{f_{0}}(r) \Theta_{f_{0}}'(r)^2 \left(r \left(\Lambda_3 r^2-1\right) \left(5 \Lambda_3 r^2-21\right) \right.\nonumber\\&&\times \left.\Theta_{f_{0}}''(r)+\left(\Lambda_3 r^2 \left(54-11 \Lambda_3 r^2\right)-63\right) \Theta_{f_{0}}'(r)\right)+16 \alpha  \Theta_{f_{0}}(r)^2 \left(r^2 \left(\Lambda_3 r^2-1\right){}^2 \Theta_{f_{0}}''(r)^2-3 \left(\Lambda_3 r^2-3\right)\right.\nonumber \\&&\times\left. \left(3 \Lambda_3 r^2-13\right) \Theta_{f_{0}}'(r)^2-2 r \left(\Lambda_3 r^2-3\right) \Theta_{f_{0}}'(r) \left(r \Theta_{f_{0}}^{(3)}(r) \left(\Lambda_3 r^2-1\right)+\left(8-4 \Lambda_3 r^2\right) \Theta_{f_{0}}''(r)\right)\right)\nonumber \\&&+\Lambda_3^4 p_{r} r^{14} \Theta_{f_{0}}(r)^4\bigg),\label{36}\\
p^{eff}_t&&=\frac{1}{\Lambda_3^4 r^{14} \Theta_{f_{0}}(r)^5}\bigg(-432 \alpha  r^3 \left(\Lambda_3 r^2-1\right) \Theta_{f_{0}}'(r)^5+4 \alpha  r^2 \Theta_{f_{0}}(r) \Theta_{f_{0}}'(r)^3 \left(208 r \left(\Lambda_3 r^2-1\right) \Theta_{f_{0}}''(r)+\left(\Lambda_3 r^2 \right.\right.\nonumber \\&&\times\left.\left.\left(9 \Lambda_3 r^2-382\right)+561\right) \Theta_{f_{0}}'(r)\right)-16 \alpha  r \Theta_{f_{0}}(r)^2 \Theta_{f_{0}}'(r) \left(16 r^2 \left(\Lambda_3 r^2-1\right) \Theta_{f_{0}}''(r)^2+\left(\Lambda_3 r^2 \left(149-3 \Lambda_3 r^2\right)\right.\right.\nonumber \\&&-\left.\left.354\right) \Theta_{f_{0}}'(r)^2+r \Theta_{f_{0}}'(r) \left(13 r \Theta_{f_{0}}^{(3)}(r) \left(\Lambda_3 r^2-1\right)+3 \left(\Lambda_3 r^2 \left(\Lambda_3 r^2-43\right)+64\right) \Theta_{f_{0}}''(r)\right)\right)+16 \alpha  \Theta_{f_{0}}(r)^3 \nonumber \\&&\times\left(\left(\Lambda_3 r^2 \left(\Lambda_3 r^2-96\right)+471\right) \Theta_{f_{0}}'(r)^2+r^2 \Theta_{f_{0}}''(r) \left(2 r \Theta_{f_{0}}^{(3)}(r) \left(\Lambda_3 r^2-1\right)+\left(\Lambda_3 r^2 \left(\Lambda_3 r^2-12\right)+19\right)\right.\right.\nonumber \\&&\times\left.\left. \Theta_{f_{0}}''(r)\right)-2 r \Theta_{f_{0}}'(r) \left(\left(\Lambda_3 r^2 \left(\Lambda_3 r^2-54\right)+135\right) \Theta_{f_{0}}''(r)+r \left(r \Theta_{f_{0}}^{(4)}(r) \left(1-\Lambda_3 r^2\right)\right.\right.\right.\nonumber \\&&+\left.\left.\left.6 \Theta_{f_{0}}^{(3)}(r) \left(2 \Lambda_3 r^2-3\right)\right)\right)\right)+\Lambda_3^4 p_{t} r^{14} \Theta_{f_{0}}(r)^5\bigg).\label{37}
\end{eqnarray}
It is interesting to notice that Eqs. (\ref{35}-\ref{37}) now involve only one unknown $\Theta_{f_{0}}(r)$.
By using the Eq. (\ref{33}) in Eq. (\ref{5.22}), we have the following differential equation
\begin{eqnarray}
&&\frac{4 \sqrt{\frac{\Lambda _2 ^2}{\Theta_{f_{0}} (r)}}}{r^4 \Theta_{f_{0}}^4 (r) \left(\Lambda_3 r^2\right){}^{7/2}}\bigg(-4 r^2 \Theta_{f_{0}} (r) \left(\Lambda_3 r^2-1\right) (\Theta_{f_{0}} '(r)^2) \left(5 r \left(\Lambda_3 r^2-1\right) \Theta_{f_{0}} ''(r)+\left(63-59 \Lambda_3 r^2\right) \Theta_{f_{0}} '(r)\right)\nonumber\\&&+ \Theta (r)^3 \left(\Lambda_3 r^2-1\right) \left(\left(5 \Lambda_3 r^2-21\right) \Theta_{f_{0}} '(r)+r \left(r \Theta_{f_{0}} ^{(3)}(r) \left(\Lambda_3 r^2-1\right)+\left(9-5 \Lambda_3 r^2\right) \Theta_{f_{0}} ''(r)\right)\right)+4 r \Theta_{f_{0}} (r)^2\nonumber\\&&\times \left(-r^2 \left(\Lambda_3 r^2-1\right){}^2 \Theta ''(r)^2+\left(\Lambda_3 r^2 \left(141 \Lambda_3 r^2-394\right)+249\right) (\Theta_{f_{0}} '(r))^2+2 r \left(\Lambda_3 r^2-1\right) \Theta_{f_{0}} '(r) \left(r \Theta_{f_{0}} ^{(3)}(r) \right.\right.\nonumber\\&&\times\left.\left.\left(\Lambda_3 r^2-1\right)\left(42-40 \Lambda_3 r^2\right) \Theta_{f_{0}} ''(r)\right)\right)+15 r^3 \left(\Lambda_3 r^2-1\right){}^2 (\Theta_{f_{0}} '(r))^4\bigg)-\mathcal{I}_{1}=0.\label{38}
\end{eqnarray}
In the context of the current study, this differential equation is the most relevant, since all physical effects depend on its solution. Since it is extremely non-linear, we can solve it numerically by computing $\Theta_{f {0}}(r)$ for some appropriate initial condition.
\begin{figure}
\centering \epsfig{file=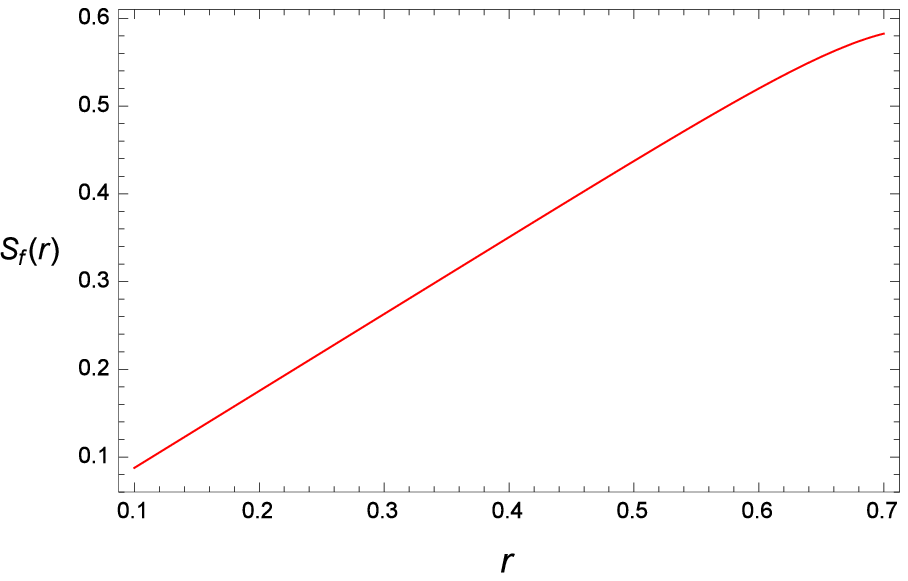, width=.45\linewidth,
height=2in}\epsfig{file=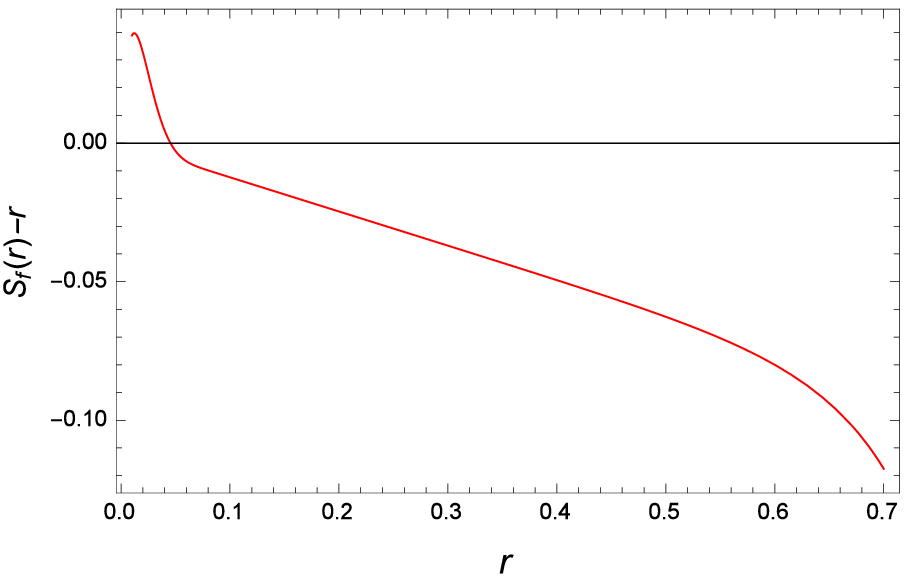, width=.45\linewidth,
height=2in} \caption{\label{fig1} is showing the behavior of $\mathbb{S}_{f}(r)$ and $\mathbb{S}_{f}(r)-r$.}
\end{figure}
\begin{figure}
\centering \epsfig{file=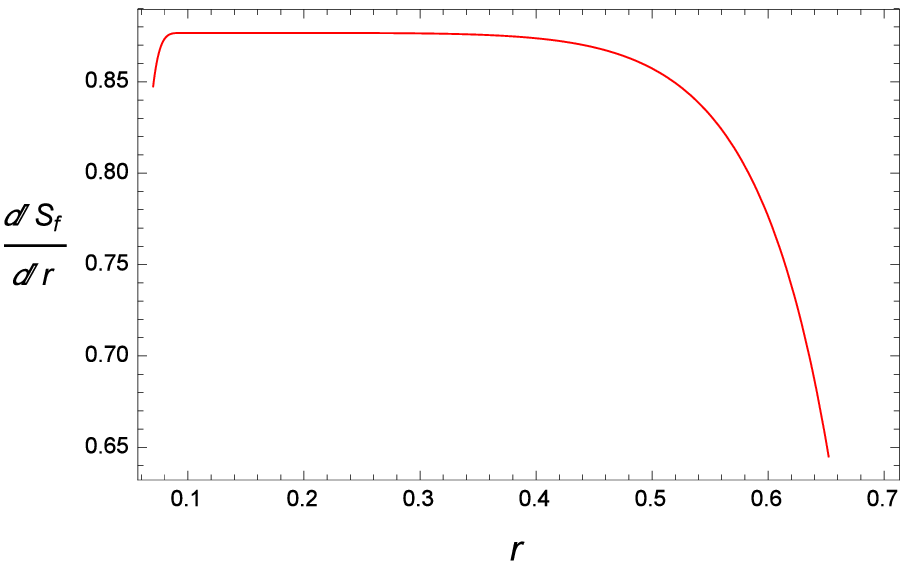, width=.45\linewidth,
height=2in}\epsfig{file=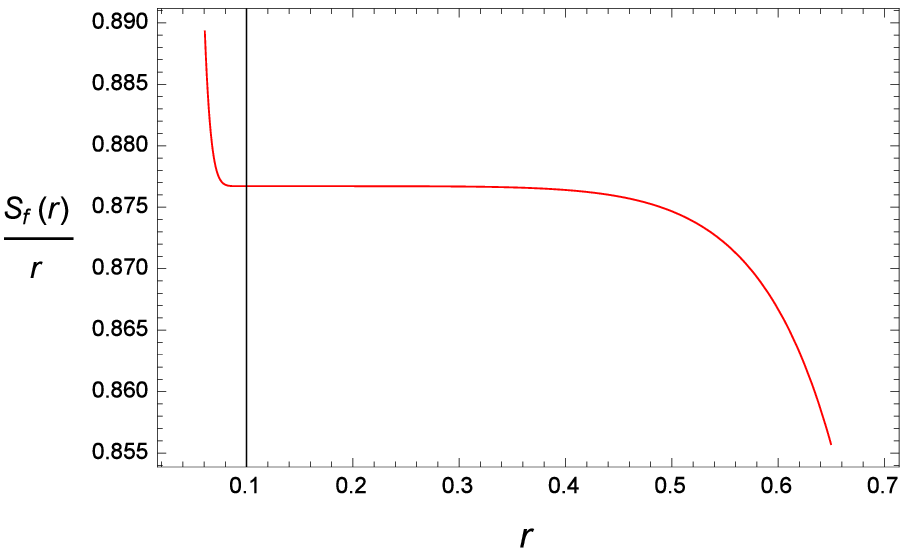, width=.45\linewidth,
height=2in} \caption{\label{fig2} is showing the behavior of $\mathbb{S}_{f}^{'}(r)$ and $\frac{\mathbb{S}_{f}}{(r)}$.}
\end{figure}
Some important results using numerical solution of Eq. (\ref{38}) are itemized below.
\begin{itemize}
\item The positive and increasing behavior of $\mathbb{S}_{f}(r)$ can be verified by the Fig. (\textbf{1}) of left panel, which justifies the existence of wormholes. The right side of the Fig. (\textbf{1}) describe the difference of shape-function and radial coordinate, i.e., $\mathbb{S}_{f}(r)-r$. Basically, this difference indicates the wormhole throat location, i.e., $r_{0}$. In particular, it can be seen from Fig. (\textbf{1}) that the difference cut the $x$ axis at $r=r_{0}=0.0420$ which provides the evidence for the opening of wormhole throat. The wormhole throat opening properties are the fundamental requirement of the wormhole geometry.

\item    The derivative of shape function with respect to radial coordinate $r$ can be perceived from the left panel of Fig. (\textbf{2}), which is seen less than one, i.e., $\frac{d\mathbb{S}_{f}}{dr}<1$ at $r_{0}=0.0420$. The condition $\frac{d\mathbb{S}_{f}}{dr}<1$ provides the flaring out property of wormhole physics. From Fig. (\textbf{2}), the ratio of shape function and radial coordinate, i.e., $\frac{\mathbb{S}_{f}(r)}{r}$ is seen by right part. In this study, the expression $\frac{\mathbb{S}_{f}(r)}{r}$ approaches to small value, i.e., $0.855$ nearing zero as $r$ approaches infinity, perhaps due to the use of Noether and conformal symmetries.
\end{itemize}

\subsection{Energy Bounds}

The energy bounds are very important for the physical acceptability of wormhole geometry. It would be an interesting task to derive modified energy bounds in frame of $f(\mathcal{G})$ gravity. For present study, we may use the $\mathcal{GR}$ energy condition with the following justification.
\\\\$\mathbf{\textit{\textbf{Theorem}:}}$
For a solution of Eqs. (\ref{10})-(\ref{12}), expressed by functions $K_1=\big\{\nu(r), \lambda(r), f(\mathcal{G})\big\}$, if we have a solution in $\mathcal{GR}$ defined by function $K_2=\big\{\nu(r), \lambda(r)\big\}$, then the energy conditions would be the same for $K_1$
and $K_2$ since $T^{eff}_{\xi\eta}$ in (\ref{5.4eff}) performs the significant role of stress energy tensor in $\mathcal{GR}$ \cite{M.V, Shamir1}.
\\\\
The energy conditions are described as null energy condition $($NEC$)$, weak energy condition $($WEC$)$, dominant energy condition $($DEC$)$, and strong energy condition $($SEC$)$, which are mentioned as \cite{Bamba}.
\begin{eqnarray}
&&NEC:  \;\;\;\forall i,\;\;\;\;\;\rho^{eff}+p^{eff}_{i}\geq 0. \;\;\;\;\;\;\;\;\;\;\;\;\;\;\;WEC:  \;\;\rho^{eff}\geq 0\;\;\;\;\;\forall i,\;\;\rho^{eff}+p^{eff}_{i}\geq 0.\nonumber\\
&&DEC:  \;\;\rho^{eff}\geq 0 \;\;\;\forall i,\;\;\;\;\;\rho^{eff} \pm p^{eff}_{i}\geq 0.\;\;\;\;\; SEC:  \;\;\rho^{eff}+p^{eff}_{i}\geq 0\;\; and \;\;\;\forall i,\;\;\;\;\;\rho^{eff}+\sum p^{eff}_{i}\geq 0\nonumber.
\end{eqnarray}
The $NEC$ and $WEC$ violation lead to exotic matter. The exotic matter is the main cause for the validity of the existence of wormhole geometry. The pictorial representation of $NEC$ and $WEC$ can be seen from the Fig. (\textbf{3}) and Fig. (\textbf{4}) respectively. We can see the violation of $NEC$ from the Fig. (\textbf{3}), which describes the presence of exotic matter. Moreover, Figs. (\textbf{5}) and (\textbf{6}) depict the evolution of radial and tangential equation of parameter. The magnified views (right panel) show that the values goes in negative range near the throat indicting existence of exotic matter. The presence of exotic matter shows the goodness and superiority of our study based on the symmetries. Due to presence of exotic matter wormhole throat should be open, which is necessary for the physically acceptable wormhole geometry.
However, a little drawback is witnessed that the physical parameters are seen justified in a narrow space due to the use of symmetry approach. It is evident from graphs that fluctuated behavior is obtained for large scale, however, magnified views show that physically viable wormholes are possible.

\begin{figure}
\centering \epsfig{file=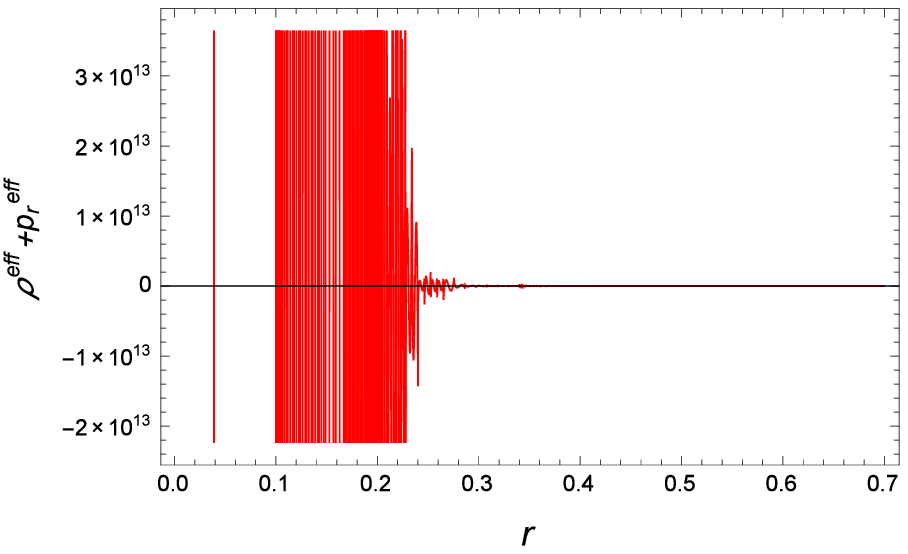, width=.45\linewidth,
height=2in}\epsfig{file=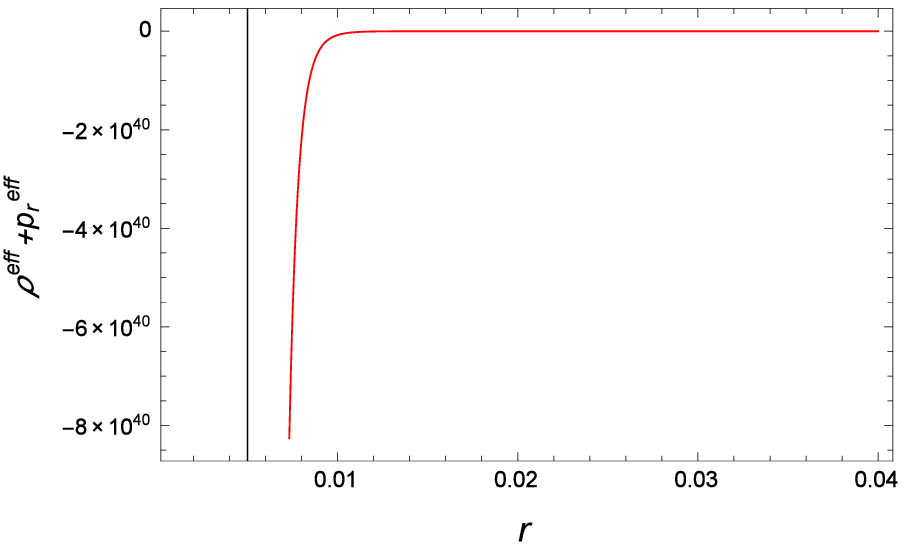, width=.45\linewidth,
height=2in} \caption{\label{fig3} is indicating the graphically behavior of $NEC$.}
\end{figure}
\begin{figure}
\centering \epsfig{file=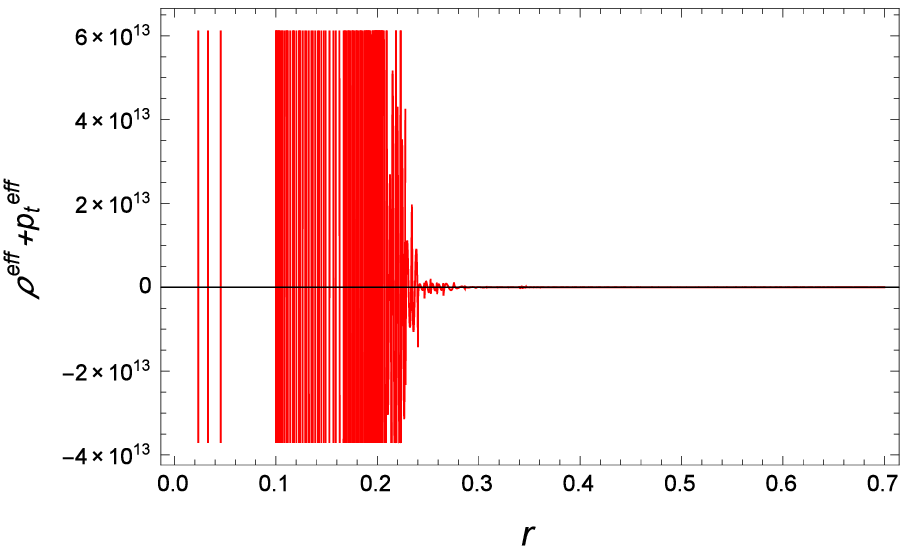, width=.45\linewidth,
height=2in}\epsfig{file=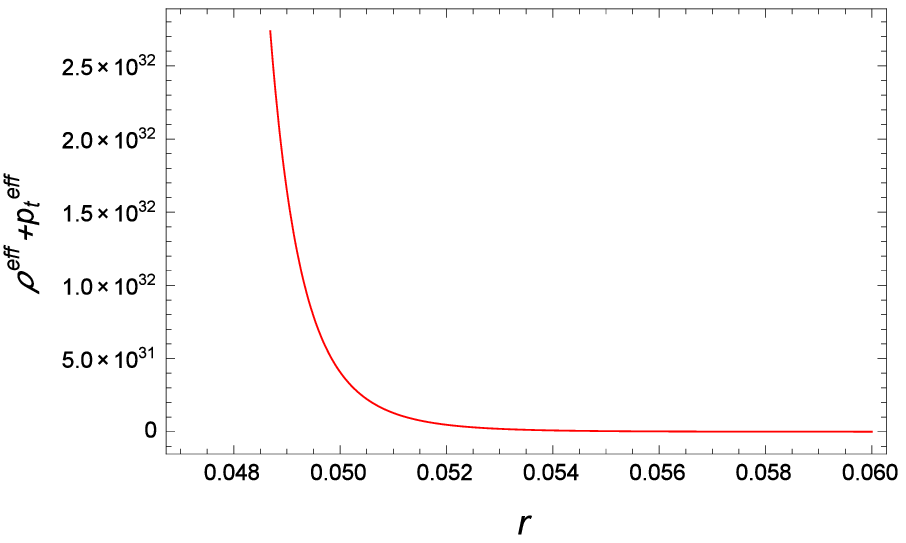, width=.45\linewidth,
height=2in} \caption{\label{fig3} is indicating the graphically behavior of $WEC$.}
\end{figure}

\subsection{Equilibrium Condition}
\begin{figure}
\centering \epsfig{file=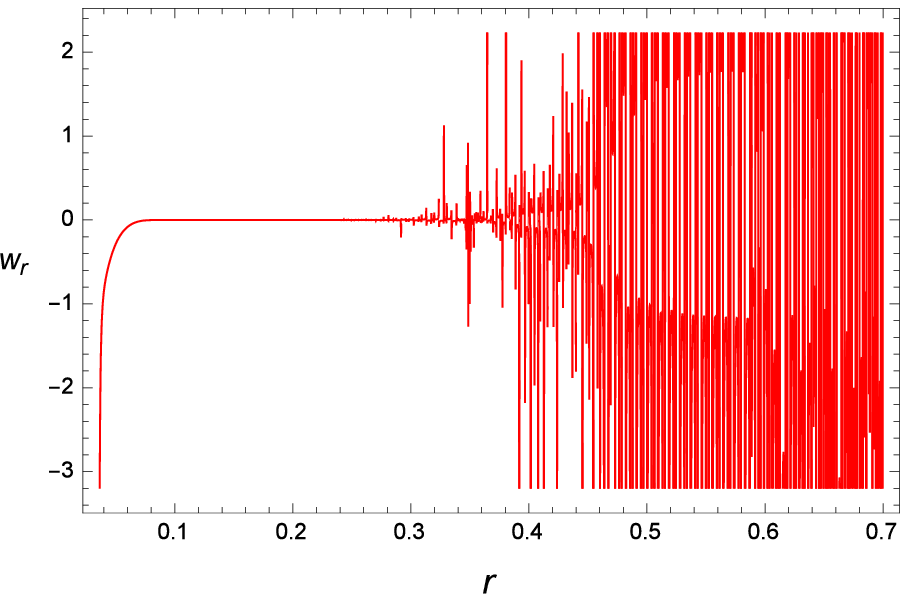, width=.45\linewidth,
height=2in}\epsfig{file=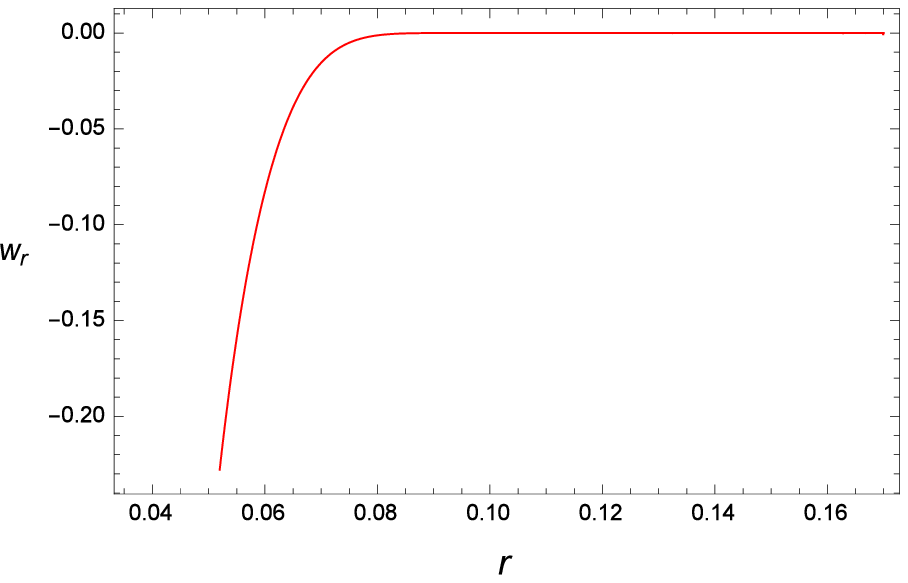, width=.45\linewidth,
height=2in} \caption{\label{fig5} shows the evolution of radial EoS parameter}
\end{figure}
\begin{figure}
\centering \epsfig{file=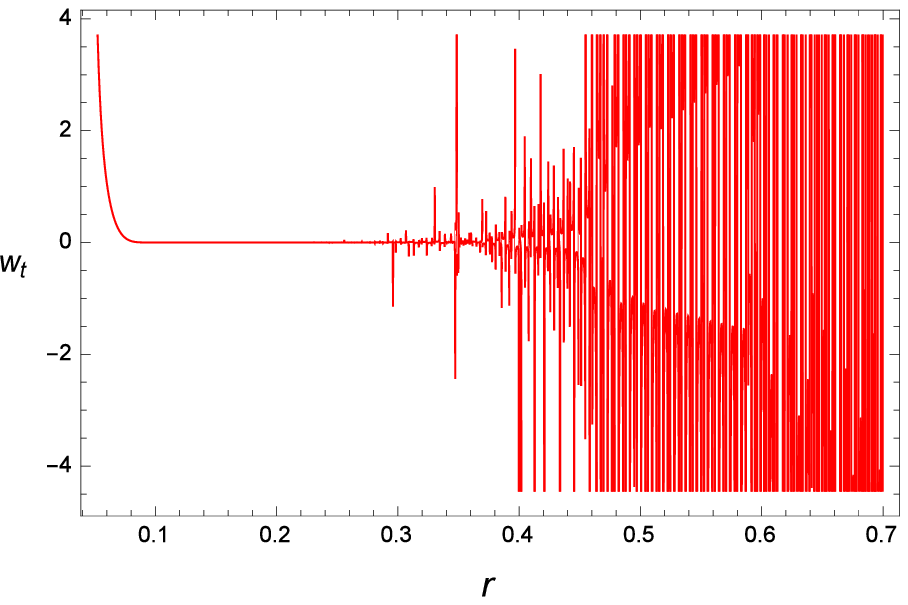, width=.45\linewidth,
height=2in}\epsfig{file=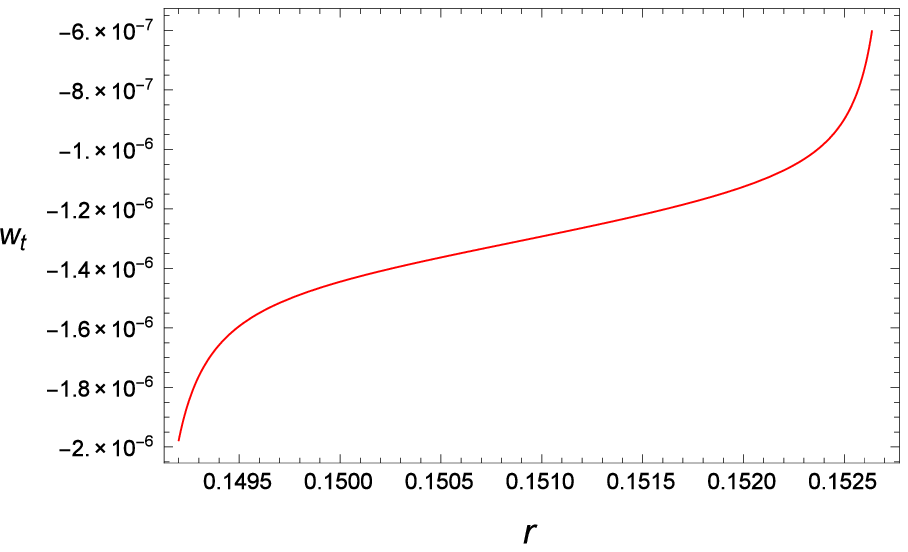, width=.45\linewidth,
height=2in} \caption{\label{fig6} shows the evolution of tangential EoS parameter}
\end{figure}
Here, we provide the stability analysis of the wormhole solutions just discussed in the previous section by incorporating the equilibrium limitations under Noether symmetry  framework with conformal symmetry, for the $f(\mathcal{G})$ gravity models under investigation. Here, we will assume Tolman-Oppenheimer-Volkoff equation, which is given as:
\begin{equation}\label{39}
-\frac{dp^{eff}_{r}}{dr}-\frac{\nu^{'}(r)}{2}(\rho^{eff}+p^{eff}_{r})+\frac{2}{r}(p^{eff}_{t}-p^{eff}_{r})=0,
\end{equation}
where $\nu(r)=2\Omega(r)$. The forces namely, hydrostatic $(\mathcal{F}_{\mathrm{h}})$, the gravitational ($\mathcal{F}_{\mathrm{g}})$ and anisotropic force $(\mathcal{F}_{\mathrm{a}})$ are represented by following expressions
\begin{equation}\label{40}
\mathcal{F}_{\mathrm{h}}=-\frac{dp^{eff}_{r}}{dr},\;\;\;\;\;\;\;\;\mathcal{F}_{\mathrm{a}}=\frac{2}{r}(p^{eff}_{t}-p^{eff}_{r}), \;\;\;\;\;\;\;\;\mathcal{F}_{\mathrm{g}}=-\frac{\nu^{'}}{2}(\rho^{eff}+p^{eff}_{r}),
\end{equation}
and thus Eq. (\ref{39}) takes the form given by
\begin{equation*}
\mathcal{F}_{\mathrm{a}}+\mathcal{F}_{\mathrm{g}}+\mathcal{F}_{\mathrm{h}}=0.
\end{equation*}
\begin{figure}
\centering \epsfig{file=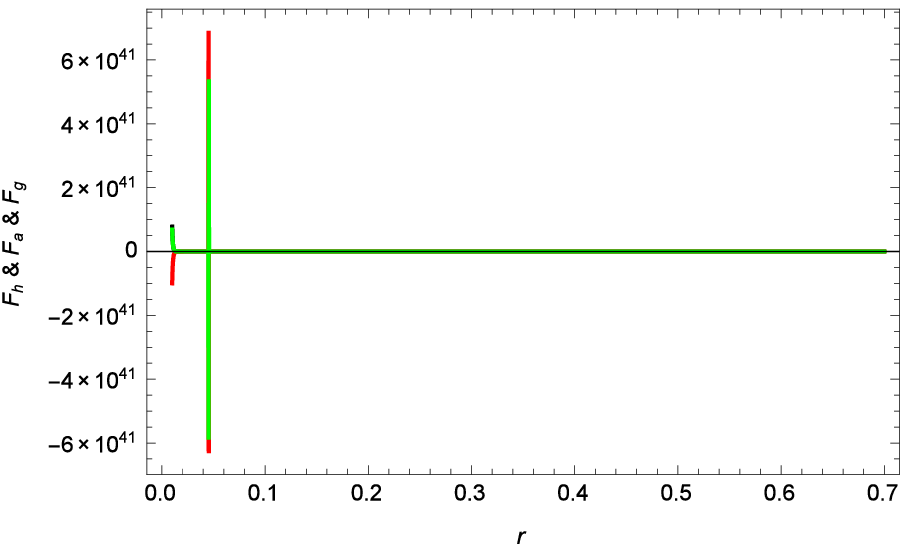, width=.45\linewidth,
height=2in}\epsfig{file=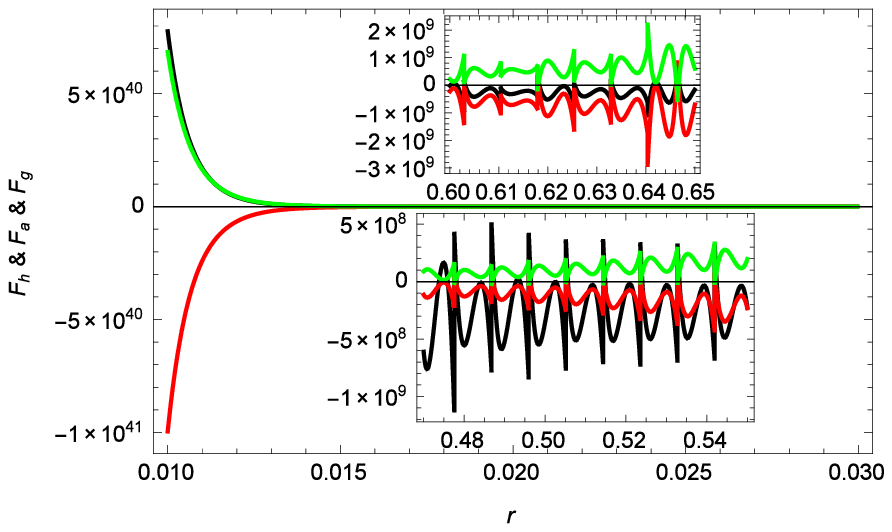, width=.45\linewidth,
height=2in} \caption{\label{fig4} indicates the balancing behavior of hydrostatic, the gravitational and anisotropic forces.}
\end{figure}
Fig. (\textbf{7}) depicts the graphical behavior of these forces, which are shown balance to each other by left part. In Fig. (\textbf{7}), the red, black, and green lines represent the hydrostatic, gravitational, and  anisotropic forces. The balancing behavior of gravitational and anisotropic forces against hydrostatic force shows the stability of wormhole existence via Tolman-Oppenheimer-Volkoff equation in the background of Noether symmetry with conformal symmetry. The right part of Fig. (\textbf{7}) shows the real impact of different forces in very small interval of radial coordinate. The stability via Tolman-Oppenheimer-Volkoff equation shows that the our inquired wormhole solutions are physically realistic and acceptable in $f(\mathcal{G})$ gravity.

\section{Conclusive Remarks}

A basic technique for solving the dynamical equations is known as the Noether symmetry technique. The Noether symmetries, in particular, are not only a mechanism for dealing with the dynamics solution, but their presence also provides suitable
conditions so that one can specify the universe models physically
and analytically according to our measured observations. The procedure of the Lagrange multiplier enables one to resolve some problems related to the $f(\mathcal{G})$ gravity model and to minimize the Lagrangian to a canonical form. It is perhaps quite useful to reduce the dynamics of the system in order to identify the exact solutions.
In this study, we examine the existence of static wormhole through Noether and conformal symmetries in the frame anisotropic matter distribution.
For this purpose, we solve overdetermined system of PDE's and find the Noether symmetry generators along with corresponding conserved quantity by incorporating the $f(\mathcal{G})$ gravity model.
Moreover, a useful conserved quantity is gained from the Noether symmetry of spherically symmetric spacetime. The presence of conserved quantity plays a significant part in defining the possible existence of wormhole solutions under Noether symmetry by utilizing conformal symmetry. Moreover, we have also investigated stable condition of wormhole solutions via modified equilibrium condition by considering the specific red-shift function. Some essential findings and observations regarding the existence of wormhole in modified $f(\mathcal{G})$ gravity are summarized below.
\begin{itemize}
\item The shape-function $\mathbb{S}_{f}(r)$ remains positive and continues to increase as shown in the left plot of Fig. (\textbf{1}). The positive behavior of shape function shows that our obtained solutions are physically acceptable.

\item  The condition $\mathbb{S}_{f}(r_0)=r_0$ is justified at $r_0=0.0420$ as depicted in Fig. (\textbf{1}), which shows that the wormhole throat should be opened.

 \item The flaring out condition, i.e., $\mathbb{S}_{f}(r)^{'}(r_{0})<1$, is satisfied, as it can be noted from the right panel of Fig. (\textbf{2}).

 \item The flatness condition also can be seen from the right panel of Fig. (\textbf{2}).

\item Referring to the energy bounds $NEC$ and $WEC$, it can be seen from the Fig. (\textbf{3}) and Fig. (\textbf{4}). The violation of $NEC$ can be perceived from the Fig. (\textbf{3}), which describes the presence of exotic matter. The presence of exotic matter shows the goodness and superiority of our study based on the symmetries, i.e., Noether symmetry and conformal symmetry. Due to presence of exotic matter wormhole throat should be open, which is necessary for the physically acceptable wormhole geometry.

\item Fig. (\textbf{7}) represents the graphical analysis of three different forces, i.e., $\mathcal{F}_{\mathrm{a}}, \mathcal{F}_{\mathrm{g}}$ and $\mathcal{F}_{\mathrm{h}}$, which are shown balance to each other by left part. In Fig. (\textbf{7}), the red, black, and green lines represent the hydrostatic, gravitational, and  anisotropic forces. The balancing behavior of gravitational and anisotropic forces against hydrostatic force shows the stability of wormhole existence via Tolman-Oppenheimer-Volkoff equation in the background of Noether symmetry with conformal symmetry.
\end{itemize}
All above discussions suggest that wormholes exist in $f(\mathcal{G})$ gravity using symmetry approach. However, a little drawback is witnessed that the physical parameters are seen justified in a narrow space due to the use of symmetry approach. It is evident from graphs that fluctuated behavior is obtained for large scale, however, magnified views show that physically viable wormholes are possible. Conclusively, it is worthy to mention here that Noether and conformal symmetries are quite helpful in obtaining physically realistic and
acceptable wormhole solutions in $f(\mathcal{G})$ gravity.\\\\

\end{document}